\documentclass{article}
\input{macro}
\usepackage{setspace}

\title{\textsc{Towards Generalizing Inferences\\ from Trials to Target Populations}\footnote{\textit{\textbf{Co-first authors; Both authors contributed equally to the paper}.\\The authors would like to thank Sarah E Robertson, Elizabeth Stuart, Kara Rudolph, Quinn Lanners for their comments and feedback on the paper. Also, the authors would like to thank the ICERM workshop organizers Elizabeth Stuart, Jon Steingrimsson, and Issa Dahabreh as well as the participants for engaging in the discourse and discussion. }}}
\author{Melody Y. Huang \thanks{HDSI Postdoctoral Fellow, Harvard University, Dept. of Statistics \& IQSS, Email: \texttt{melodyhuang@fas.harvard.edu}} \quad Harsh Parikh \thanks{Postdoctoral Fellow, Johns Hopkins University, Dept. of Biostatistics, Email: \texttt{hparikh4@jh.edu}}}

\begin{document}

\maketitle

\begin{abstract}
Randomized Controlled Trials (RCTs) are pivotal in generating internally valid estimates with minimal assumptions, serving as a cornerstone for researchers dedicated to advancing causal inference methods. However, extending these findings beyond the experimental cohort to achieve externally valid estimates is crucial for broader scientific inquiry. This paper delves into the forefront of addressing these external validity challenges, encapsulating the essence of a multidisciplinary workshop held at the Institute for Computational and Experimental Research in Mathematics (ICERM), Brown University, in Fall 2023. The workshop congregated experts from diverse fields including social science, medicine, public health, statistics, computer science, and education, to tackle the unique obstacles each discipline faces in extrapolating experimental findings. Our study presents three key contributions: we integrate ongoing efforts, highlighting methodological synergies across fields; provide an exhaustive review of generalizability and transportability based on the workshop's discourse; and identify persistent hurdles while suggesting avenues for future research. By doing so, this paper aims to enhance the collective understanding of the generalizability and transportability of causal effects, fostering cross-disciplinary collaboration and offering valuable insights for researchers working on refining and applying causal inference methods.
\end{abstract}
\textbf{\textit{Keywords: }} {Causal Inference, Generalizability, External validity, Machine Learning}

\clearpage
\doublespacing 

\section{Introduction}
Randomized controlled trials (RCTs) are often considered the gold standard for estimating causal effects. From field experiments to medical clinical trials, RCTs form the bedrock of decision-making in social sciences, public health, medicine, and other fields. However, in practice, researchers generally care not only about the estimated effect within their experimental sample, but also about the impact of a specific intervention beyond the experiment, within a larger, or different target population. Due to feasibility or ethical constraints, RCTs rarely comprise a representative sample from the target population, limiting the \textit{external validity} of the estimated results. 

Limitations to external validity arise from differences between the trial and target populations. These differences can be a result of (i) observed or unobserved pretreatment characteristics, (ii) settings such as geography, the timing of treatment, dosage, or staff training, and (iii) outcomes, such as the length and timings of measurements \citep{degtiar2023review}. Challenges to the generalizability of trials prompt careful consideration of necessary and realistic assumptions for the identification and estimation of causal effects for the target population \citep{stuart2015assessing}. A variety of posthoc adjustment methods have been proposed to account for differences in the trial and target populations, allowing researchers to estimate externally valid causal effects under certain assumptions \citep[e.g.,][]{ chipman2007, stuart2011, kern2016, rudolph2017, bennett2020, rudolph2023efficiently, dahabreh2019generalizing}. 

This paper provides a comprehensive overview of recent work discussed at the Fall 2023 workshop hosted by the Institute for Computational and Experimental Research in Mathematics (ICERM) at Brown University on ``Extending Inferences from Trials to Target Populations.'' The workshop brought together researchers from diverse interdisciplinary backgrounds, including social science, medicine, public health, statistics, computer science, and education. Each discipline's unique settings and challenges introduce distinct issues when generalizing experimental results. For example, in the context of trials, ethical concerns about potential comorbidities from treatment and preventing negative outcomes are paramount in medicine and public health, leading to selective criteria in trial participation. This can make it difficult to assess the effectiveness of medical interventions on individuals systematically excluded from trials. In education, researchers often face challenges from small sample sizes and the complexities of clustered randomized controlled trial settings, making it crucial to understand treatment effect variation with limited units to design externally valid studies.


In what follows, first, we identify three major themes that encompass the presentations at the workshop: (i) assessment of the external validity of trials (Section~\ref{sec: external_val}), (ii) considering external validity beyond intention-to-treat effects (Section~\ref{sec: internal_val}), and (iii) leveraging advancements in machine learning approaches to improve estimation in external validity (Section~\ref{sec: ml}). 
For each section, we contextualize these advancements within the existing literature. This is followed by an in-depth discussion on recent advancements presented at the workshop, organized into several broader sub-themes for each main theme. Finally, based on the discussions at the workshop, we conclude the paper with a discussion on five key future research directions and briefly introduce the corresponding challenges and opportunities in Section~\ref{sec: conclude}. We provide a list of talk titles and speakers in the Appendix.

Throughout the paper, we will refer to \textit{generalizability analyses} as interest in learning about a restricted target population that meets all trial eligibility criteria and \textit{transportability analyses} as interest in learning about a target population broader than the trial eligibility criteria \citep{dahabreh2019extending}. In particular, we refer to \textit{transportability} as the setting in which the trial population is a non-representative subset of the target population.

\section{Validity of Identifying Assumptions under Covariate Shifts}\label{sec: external_val}
\textbf{Context.} One primary source of bias when aiming to estimate externally valid causal effects arises from differences in the underlying distribution of treatment effect moderators between the experimental sample and the target population \citep[e.g.,][]{imai2008misunderstandings, cole2010generalizing, olsen2013external, egami2022elements}. Informally, moderators are covariates that describe how the treatment will differentially affect individuals. Thus, to estimate a valid causal effect across a target population of interest, researchers must adjust for the covariate shift in the underlying moderators between the experimental sample and the target population. Two common identifying assumptions leveraged in practice are (1) selection on observables (also referred to as conditional exchangeability) and (2) positivity of trial participation.

The first assumption, selection on observables (also referred to as conditional ignorability of treatment effect heterogeneity and selection, or conditional mean exchangeability \citep[e.g.,][]{dahabreh2019generalizing, cole2010generalizing, olsen2013external}), assumes that researchers can adjust for all distributional differences in moderators between the experimental sample and the target population. However, this is often difficult to justify. Researchers frequently lack comprehensive covariate data for the target population, rely on inconsistent data sources like electronic health records or surveys, and may face measurement differences and the challenge of identifying all relevant moderators. The second assumption, the positivity of trial participation, requires that all units in the target population have a non-zero chance of being included in the experimental sample, but this can be violated due to logistical, financial, and ethical constraints. Logistical and financial priorities often focus on maximizing power and efficient recruitment rather than external validity, while ethical considerations can lead to exclusion criteria that omit subsets of the target population, such as pregnant individuals in early-stage medical trials.

While the combination of these assumptions allows researchers to theoretically identify the target population's average treatment effect, in practice, they can be difficult to justify. Several recent advancements focus on methods to help researchers consider the validity of the underlying identifying assumptions. We summarize key themes below: \\ \\ 
(1) \textbf{Understanding Treatment Effect Heterogeneity.} Both selection on observables and positivity of trial participation rely on researchers knowing which covariates moderate the treatment effect of the intervention of interest. This is important for both the design stage of an experiment--i.e., understanding which individuals to recruit into the experimental sample--and for post-hoc adjustments--i.e., knowing which covariates to collect across both the experimental sample and the target population. One of the presentations at ICERM, \citet{tipton2023designing}, provides a new view on the importance of understanding the underlying moderators by formulating the challenge of generalizing experimental results as a prediction problem. In particular, the authors highlight that in the absence of a sufficiently heterogeneous experimental sample, the target populations that the experimental results can be generalized (or transported) to will be limited. Furthermore, not only will standard estimated effects be biased for population quantities, but the associated uncertainty will also be incorrect, making inference challenging as well. In the absence of knowledge about heterogeneity, it can be challenging to effectively recruit a representative experimental sample, and it will be infeasible to perform posthoc adjustment.  \\ \\ 
(2) \textbf{Developing Methods for Researchers to Incorporate Substantive Knowledge.} Incorporating substantive knowledge into research methodologies is crucial for formulating effective policy recommendations, especially when utilizing existing experimental data in new domains. Two of the talks at ICERM focused on this. The first talk, \citet{gechter2018evaluating}, introduces a decision-theoretic approach to integrating expert knowledge with ex-post experimental evidence for assessing modeling choices for policy recommendation. The paper proposes a methodological framework for researchers to customize the utility of a decision-maker into evaluating which models are preferred in generating counterfactual predictions. The utility function can accommodate non-linear preferences, as well as the cost of treatments. Using the proposed framework, researchers can then compare different approaches to prediction, using the specified, custom preferences encoded in the welfare function. This process aids in selecting the appropriate method for generating preliminary recommendations in contexts lacking experimental data. Similarly, addressing the challenge of positivity violations—where certain subgroups are absent from the sample—requires reliance on substantive knowledge or mechanistic models. The second talk, \citet{zivich2023synthesis}, introduces a novel approach that employs deterministic mathematical models to calculate treatment effects for segments of the target population not included in the experimental sample. This method then integrates these calculations with traditional statistical models for areas where data overlap exists, providing a comprehensive estimation of the average treatment effect for the entire target population. \\ \\ 
(3) \textbf{Assessing Validity of Identifying Assumptions.} Sensitivity analyses allow researchers to consider the robustness of results to violations of the underlying assumptions. In the context of evaluating selection on observables assumptions, sensitivity analyses often take the form of parameterizing the bias that arises from omitting a moderator, either from the estimated weights \citep[e.g.,][]{cole2010generalizing, stuart2011, buchanan2018generalizing}, or the underlying treatment effect heterogeneity model \citep[e.g.,][]{kern2016}. If researchers find that even for a relatively weak moderator, the resulting bias is very large, then this implies that there is a relatively large degree of sensitivity to an omitted moderator. In contrast, if researchers find that the bias is small even for relatively strong moderators, then this implies a greater degree of robustness. One of the talks at ICERM, \citet{huang2022sensitivity}, introduces a sensitivity framework for researchers to consider violations in the selection on observables assumption. In particular, \citet{huang2022sensitivity} decomposes the bias from omitting a moderator (i.e., a violation in the underlying selection on observables assumption) into two sensitivity parameters: an $R^2$ value that represents how imbalanced an omitted moderator is across the trial sample and target population and a correlation term that represents how correlated the omitted moderator is to the individual-level treatment effect. The paper also introduces a suite of sensitivity tools that allow researchers to summarize the robustness of their generalized estimates transparently and systematically. These tools include numerical summary measures, like the robustness value, visual summary measures in the form of bias contour plots, and a benchmarking approach that allows researchers to incorporate the observed covariates into the sensitivity analysis and calibrate what might be plausible sensitivity parameters.

\section{Beyond Transporting intention-to-treat Effects}\label{sec: internal_val}
\noindent \textbf{Context.} 
The intention-to-treat (ITT) effect considers the effect of assigning treatment, rather than whether the treatment is actually received. Most existing approaches in external validity have focused on generalizing or transporting ITT effects. However, variation in treatment adherence is common after randomization, significantly affecting the interpretation of trial results. Transporting ITT effects without considering heterogeneity in treatment adherence assumes that the compliance behaviors observed in the RCT are representative of those in the target population. If treatment effects vary significantly across different compliance patterns, the reported effect estimate may fail to capture the true impact of the interventions accurately. This mischaracterization can significantly obscure real treatment effects. Furthermore, in practice, researchers and policymakers may have substantive interest in not only the generalized ITT, but also generalized impacts of complier average causal effects. 

\noindent Two recent works presented and discussed at the workshop, \citet{dahabreh2022generalizing} and \citet{huling_talk_adherence}, address the challenge of transporting treatment effect under lack of total adherence by focusing on identifiability of intent-to-treat effect in the target population, defining new relevant estimands and proposing alternative trial designs: \\ \\
    (1) \textbf{Transportability of Intent-to-Treat Effect.} \citet{dahabreh2022generalizing} highlights that the effect of assignment may be unidentifiable, especially when non-adherence is present. In particular, \citet{dahabreh2022generalizing} provides two scenarios where non-adherence significantly hinders the ability to generalize inferences. The first scenario deals with a case where trial participation directly affects treatment receipt, which influences the outcome through treatment receipt, resulting in a ``trial engagement effect'' via adherence. In the second scenario, trial participation shares unmeasured common causes with treatment receipt. In both these scenarios, the effect of assignment (i.e. intent-to-treat effect) on the outcome in the target population is not identifiable. However, in the first scenario, the effect of joint interventions that scale up trial activities influencing adherence and assignment is identifiable.\\ \\ 
    (2) \textbf{Alternate Estimand and Estimation.} Under non-adherence to treatment assignment, \citet{huling_talk_adherence} as well as \citet{dahabreh2022generalizing} presents a different causal estimand that may be of interest instead, such as the compiler average treatment effect among the target population or per-protocol effect. \citet{huling_talk_adherence} discusses a principal stratification framework to identify causal effects by conditioning on both compliance behavior and membership in the target population. They also formulate non-parametric efficiency theory and construct efficient estimators for these transported principal causal effects, characterizing the finite-sample performance through simulation experiments.\\ \\ 
    (3) \textbf{Double-blind versus Pragmatic Trials.}  \citet{dahabreh2022generalizing} discusses the challenge with the assumption of ``variation irrelevance for treatment assignment'' that states that there is no direct effect of assignment on the outcome. In other words, the outcome is not affected by how the treatment is assigned. However, the assignment may include double-blinding in the trial, which will not occur in clinical practice and may directly affect the outcome. Therefore, \citet{dahabreh2022generalizing} advocates operationalizing pragmatic trials, with open-label assignment, which may be more transportable and more likely to have the assignment variation irrelevance hold. However, it is important to note that this lack of blinding can affect the behaviors of both those delivering and receiving the intervention, potentially skewing the results. Furthermore, while pragmatic trials are lauded for their high external validity, allowing for a broader patient population and real-world clinical settings, this often comes at the cost of internal validity. The less stringent controls over participant selection and intervention administration can lead to variability that makes it difficult to attribute outcomes directly to the intervention being tested.

\section{Machine Learning and Generalizability}\label{sec: ml}
\noindent \textbf{Context.} Recent work in machine learning (ML) has enabled researchers to flexibly estimate causal relationships between variables in complex data settings \citep{chernozhukov2018double, rudolph2017robust, morucci2023double, brantner2023methods, colnet2020causal, robertson2023regression}. ML methods offer several advantages over traditional approaches, particularly in their ability to flexibly handle high-dimensional data with nonlinear dependencies \citep{chernozhukov2018double, morucci2023double, huang2023leveraging}. By effectively modeling these nuisance parameters, ML algorithms can provide more accurate estimates of heterogeneous causal effects, which is crucial to understanding generalizability. The works discussed at the workshop are grouped into the following three core themes focused on leveraging recent advances in ML to improve estimation in the context of generalizability:\\ \\ 
    (1) \textbf{Estimating heterogeneous effects.} Identifying effect moderators is a key part of understanding the potential generalizability or transportability of study results. 
    With their ability to capture complex nonlinear relationships and handle high-dimensional data, ML methods are well-suited to estimate heterogeneous treatment effects. One talk at the workshop, \citet{rudolph2023efficiently}, introduces a collaborative one-step estimator for scenarios where researchers lack information about which covariates modify effects and which vary in distribution across different populations. Similarly, \citet{kim2022universal}, presented at the ICERM workshop, develops an approach to transport inferences to an (unknown) target population by leveraging the connection between statistical inference and algorithmic fairness through ``multi-calibration.'' The core idea behind multi-calibration is to reduce bias across subpopulations by ensuring that the predictor is well calibrated, not only overall but also within specific subpopulations.\\ \\ 
    (2) \textbf{Addressing Underrepresentation in RCTs.} A significant challenge in generalizing RCT findings is ensuring that the benefits of evidence-based interventions reach all segments of society, including underrepresented groups. RCTs often struggle to recruit and retain participants from diverse backgrounds, leading to a lack of data on how interventions may affect these populations. \citet{parikh23missing}, discussed at the workshop, designed a nonparametric ML procedure, Rashomon set of optimal trees (ROOT), addressing this challenge by providing interpretable characterizations of underrepresented groups in the trial population. By identifying patterns in the data that differentiate these groups, ROOT can help researchers understand the factors that contribute to their underrepresentation and develop strategies to improve recruitment and retention. Further, another work at the workshop by \citet{elliott2023improving} uses robust and flexible ML methods such as BART \citep{chipman2007} for extending inference to the broader population. This work focuses on addressing biases caused by effect modifiers and systematic differences between trial and target populations by leveraging advanced prediction and weighting techniques.\\ \\
    (3) \textbf{Leveraging Multiple Data Sources.} Another area where machine learning approaches have driven advancements is in data fusion, which enhances the generalizability, precision, and accuracy of inferences across different populations. As previously discussed by \citet{bareinboim2016causal}, while data fusion is a cost and time-efficient alternative to conducting extensive cohort RCTs, integrating heterogeneous data sources can be challenging. It requires a deep understanding of the background knowledge and new analytical tools for identification and estimation. Recent work by \citet{schnitzler2023two} presented at the ICERM workshop argues that traditional meta-analytic approaches often fail to provide causally interpretable estimates due to treatment effect heterogeneity across studies. To address this, \citet{schnitzler2023two} proposes a two-stage approach to extend inferences from multiple RCTs, allowing for interpretable estimates in a target population using a weighted average of study-specific treatment effect estimates. Similarly, ML methods can be employed to borrow information from external control datasets, even if these datasets do not represent the RCT population perfectly. Meanwhile, \citet{cheng2023enhancing}, presented at the ICERM workshop, discusses benefits and challenges while integrating external controls with the RCT data for efficiency gains. Particularly, when the concurrent controls in RCT and the external controls might be incomparable, \citet{cheng2023enhancing} introduces a robust method for estimating treatment effects by integrating external controls using a new estimator called the Double Penalty Integration Estimator (DPIE) that ensures consistency, the oracle property, and asymptotic normality. However, achieving this requires careful selection of hyperparameters that effectively handle variations across the data. Thus, while combining multiple sources of data (e.g., as in a meta-analysis of individual patient data to an external target population) can yield gains in precision and generalizability, they often require alternative untestable assumptions.

\section{Discussion and Conclusion}\label{sec: conclude}
Our paper summarizes some of the recent advancements in causal methods for generalizability and transportability, as presented at the ICERM workshop. The workshop focused on three key areas of focus: (i) evaluating the robustness of foundational assumptions, (ii) extending the scope of causal analysis beyond traditional intention-to-treat effects, and (iii) incorporating machine learning to enhance the applicability of findings across various contexts. The discussion at the workshop also highlighted several challenges in bridging the research-to-practice gap as well as interesting future research directions. We provide a few examples of key discussion points from the workshop.\\ \\ 
    (1) \textbf{Addressing Data Harmonization Challenges:} Most methods in the literature, including nearly all the work presented here, require data that is harmonized across trial and population. In other words, the data in the experiment and the target population must be measured in the same way and be encoded in the same way. Furthermore, the treatment administered in the experiment must also be administered in the same way in the target population. Constructing harmonized data poses a significant challenge and is not straightforward, especially given the current lack of coordination of measurement tools across trials and population data. A concerted effort is needed to design approaches for data harmonization, along with remedial strategies when harmonization is not feasible.\\ \\
    (2) \textbf{Addressing the Underrepresentation in Trials:} While many approaches discussed here focus on assessing violations of positivity and lack of representation, these approaches generally rely on post hoc adjustments and require strong assumptions. More work must be done to consider how to \textit{design} future studies to provide information on target populations holistically, especially when the target population comprises of underrepresented groups. \\ \\
    (3) \textbf{Understanding the Underlying Causal Mechanism:} Inferring causal effects is not merely a statistical challenge; it heavily relies on underlying domain knowledge for the careful consideration of necessary assumptions. There is a need for a strong partnership between domain experts and methodologists to deepen the understanding of the underlying causal mechanisms and the plausibility of assumptions. Furthermore, \citet{tipton2023designing} highlighted the importance of moderators in recruiting a sufficiently heterogeneous experimental sample to generalize experimental results properly; however, a critical question remains of \textit{how} researchers should determine what are important moderators \textit{a priori}. Future work is needed to help establish systematic and data-driven approaches to help researchers determine salient characteristics in their target population that may moderate the treatment effect. \\ \\
    (4) \textbf{Establishing Best Practice and Improving Interpretability:} While recent methodological developments have introduced different tools for researchers to consider transporting or generalizing their causal effects, more work is needed to establish best practices and guidance for interpretability into the data analysis pipeline. For example, as discussed in Section \ref{sec: external_val}, several of the proposed methods provided ways for researchers to integrate their substantive knowledge into assessing the validity of the underlying assumptions (i.e., \citealp{huang2022sensitivity}'s benchmarking approach, as well as \citealp{zivich2023synthesis}'s bounding method). While this provides an opportunity for researchers to leverage their substantive knowledge when considering the validity of the underlying identifying assumptions in more transparent and systematic ways, future work is needed to establish best practices and minimize potential misuse of these tools. For example, the pre-registration of moderators to benchmark could help minimize researcher bias when reporting sensitivity measures. Furthermore, many of the methodological approaches rely on complex statistical machinery that is often black-box to domain experts. Developing interpretable approaches by collaborating with domain experts would help improve the translational ability of the methods, saving time and effort while making the methods more applicable.\\ \\
    (5) \textbf{Leveraging Connection to Recent ML Advancements:} A separate but related field of research in machine learning literature, with significant overlap with causal inference, encompasses domain adaptation \citep{kouw2018introduction}, distributionally robust optimization \citep{rahimian2019distributionally}, and transfer learning \citep{pan2009survey}. These disciplines share the common challenges associated with generalizing or transporting inferences\footnote{See Appendix for a brief note on the connections between causal inference and these domains}. However, the lack of a shared vocabulary has thus far limited cross-disciplinary collaboration. A focused effort is needed to bridge the linguistic and research gaps between these areas, which would advance the development of new methodologies.

These directions are pivotal for designing approaches that are not only well-grounded and accurate but also practical in addressing real-world, high-stakes issues in generalizability and transportability. By bringing together the expertise of domain specialists and methodological rigor, these collaborations ensure the reliability and applicability of the developed approaches. 

\bibliographystyle{apalike}
\bibliography{references}

\clearpage
\section*{Appendix}
\noindent \textbf{Connections to Domain Adaptation, Distributionally Robust Optimization, Transfer Learning:} In these areas, the training data used for fitting the model differs from the target dataset of interest, often unknown beforehand. In domain adaptation, the goal is to adapt a model trained on one domain to perform well on a different domain \citep{redko2019advances}. This is often relevant in causal inference settings where the trial cohort may not perfectly represent the target population. Distributionally robust optimization aims to develop models that are robust to changes in the distribution of the data, ensuring that their performance remains consistent across different target populations \citep{duchi2018learning,rahimian2019distributionally,staib2019distributionally}. Transfer learning focuses on transferring knowledge from a source task to a target task, even with different input or output spaces \citep{pan2009survey,zhuang2020comprehensive}. This approach is beneficial in scenarios where the target dataset is small or limited, allowing researchers to leverage insights from a related task to improve performance on the target task.\\

\noindent \textbf{Workshop Talk Titles and Authors}
\begin{enumerate}
    \item Evaluating Ex Ante Counterfactual Predictions Using Ex Post Causal Inference \citep{gechter2018evaluating}\\
    Speaker: Cyrus Samii
    \item {Efficiently transporting average treatment effects using a sufficient subset of effect modifiers} \citep{rudolph2023efficiently}\\
    Speaker: Kara Rudolph
    \item {Understanding effect heterogeneity in observational and randomized studies of causality} \citep{vo2022heterogeneity}
    \\
    Speaker: Ivan Diaz
    \item Generalizing trial evidence to target populations in non-nested designs: Applications to AIDS clinical trials \citep{li2022generalizing}\\
    Speaker: Ashley Buchanan
    \item {Extending Inferences to a Target Population Without Positivity} \citep{zivich2023synthesis}\\
    Speaker: Paul Zivich
    \item {Generalizing and transporting inferences about the effects of the treatment assignment subject to non-adherence} \citep{dahabreh2022generalizing}\\
    Speaker: Sarah Robertson
    \item {Who Are You Missing?: A Principled Approach to Characterizing the Underrepresented Population} \citep{parikh23missing}\\
    Speaker: Harsh Parikh
    \item Using external data to address measurement error: a transportability perspective \citep{ross2024leveraging}\\
    Speaker: Rachael Ross
    \item {Mitigating Bias in Treatment Effect Estimation: Strategies for Utilizing External Controls in Randomized Trials} \citep{cheng2023enhancing}\\
    Speaker: Shu Yang
    \item {Improving Transportability of Randomized Clinical Trial Inference Using Robust Prediction Methods} \citep{elliott2023improving}\\
    Speaker: Michael Elliott
    \item {Universal adaptability} \citep{kim2022universal}\\
    Speaker: Frauke Kreuter
    \item {Sensitivity Analysis for Generalizing Experimental Results} \citep{huang2022sensitivity}\\
    Speaker: Melody Huang
    \item {Beyond Generalization: Designing Randomized Experiments to Predict Treatment Effects} \citep{tipton2023designing}\\
    Speaker: Elizabeth Tipton
    \item Extending Inferences from a Diverse Collection of Trials \citep{schnitzler2023two}\\
    Speaker: Eloise Kaizar
    \item Transportability of Causal Effects in Principal Strata \citep{huling_talk_adherence}\\
    Speaker: Jared Huling
\end{enumerate}

\end{document}